\documentclass[12pt]{iopart}
\usepackage{iopams}
\usepackage[T1]{fontenc}
\usepackage[latin9]{inputenc}
\usepackage{graphicx}
\usepackage{amssymb}
\usepackage{amsmath}
\usepackage[english]{babel}
\begin{document}

\title{Clustered Chimera States in Systems of Type-I Excitability}

\author{Andrea V\"ullings$^1$, Johanne Hizanidis$^2$, Iryna Omelchenko$^{1,3}$, Philipp H\"ovel$^{1,3}$}
\address{$^1$Institut f{\"u}r Theoretische Physik, Technische Universit\"at Berlin, Hardenbergstra\ss{}e 36, 10623 
Berlin, Germany}
\address{$^2$National Center for Scientific Research ``Demokritos'', 15310 Athens, Greece}
\address{$^3$Bernstein Center for Computational Neuroscience, Humboldt-Universit{\"a}t zu Berlin, Philippstraße 13, 10115 Berlin, Germany}
\ead{ioanna@chem.demokritos.gr}

\begin{abstract}
Chimera is a fascinating phenomenon of coexisting synchronized and
desynchronized behaviour that was discovered in networks
of nonlocally coupled identical phase oscillators over ten years ago.
Since then, chimeras were found in numerous theoretical
 and experimental studies 
and more recently in models of neuronal dynamics as well.
In this work, we consider a generic model for a saddle-node bifurcation on a limit cycle representative for neural 
excitability type I.
We obtain chimera states with multiple coherent regions (clustered chimeras/multi-chimeras)
depending on the distance from the excitability threshold, the range of
nonlocal coupling as well as the coupling strength. A detailed stability diagram for these chimera states as well as
other interesting coexisting patterns like traveling waves are presented.
\end{abstract}

% Bifurcation nonlinear dynamics, 05.45.-a
% neuronal network dynamics, 87.19.lj
% Synchronization, nonlinear dynamics, 05.45.Xt
% Pattern formation in complex systems, 89.75.Kd
\vspace{2pc}
\submitto{\NJP}

\maketitle

\section{Introduction}
About ten years ago, a peculiar synchronization phenomenon was reported: in a system of nonlocally coupled oscillators, 
a state was discovered where synchronous and asynchronous oscillators 
coexist, even though the oscillators are identical and the interaction symmetric and translational 
invariant \cite{KUR02a}.
This phenomenon was termed the name ``chimera''  after the monstrous fire-breathing creature of Greek mythology 
composed of the parts of three different animals, a lion, a snake and a goat \cite{ABR04}. From the perspective of 
nonlinear dynamics, this surprising break of symmetry is observed by the coexistence of incongruent states of spatial 
coherence and disorder.

Real-world examples that exhibit a chimera state include electric-power grids, which rely on synchronized generators 
to avoid blackouts in power transmission. Also, certain patterns of intense heart-tissue contraction known as ``spiral 
waves'' in certain types of heart attacks have been observed in simulations of chimera states. Forms of chimera state 
may also be connected to large-scale synchronization patterns of neurons that have been observed during seizures. For a 
comprehensive review refer to \cite{PAN14} and references therein.

Chimera states were first reported by Kuramoto and Battogtokh in a model of densely and uniformly distributed 
oscillators, described by the complex Ginzburg-Landau equation in one spatial dimension, with nonlocal coupling of 
exponential form \cite{KUR02a}. This seminal work was followed by the work of Abrams and Strogatz \cite{ABR04}, who 
observed this phenomenon in a 1-dimensional ring continuum of phase oscillators assuming nonlocal coupling with a 
cosine kernel and coined the word  ``chimera'' for it. The same authors also found chimera states in networks of 
identical, symmetrically coupled Kuramoto phase oscillators \cite{ABR08} by considering two subnetworks with 
all-to-all coupling both within and between subnetworks, assuming strong coupling within each subnetwork and weaker 
coupling between them. This coupling scenario was also employed by C.~G. Laing who demonstrated the presence of chimeras in coupled 
Stuart-Landau oscillators \cite{LAI10}.
More recently, the same coupling scheme was used in a system of pendulum-like elements represented by phase oscillators
with a second derivative term, where chimera states were also investigated~\cite{BOU14}.
Furthermore, Stuart-Landau oscillators have also been investigated related to amplitude-mediated chimera \cite{SET13} and for symmetry-breaking coupling. The latter leads to a combination of chimeras and oscillation suppression, termed chimera death \cite{ZAK14}.
Chimeras have also been observed in many other systems, including coupled chaotic logistic maps and R\"ossler models 
\cite{OME11,OME12}. 
Together with numerical, the theoretical studies of chimera states have been recently provided, such as general
bifurcation analysis for chimeras with one and multiple incoherent domains in the system of nonlocally coupled phase oscillators~\cite{OME13a}.

The first experimental evidence of chimera states was found in populations of coupled chemical oscillators as well as in optical coupled-map lattices realized by liquid-crystal light modulators \cite{TIN12,HAG12}.
Recently, Martens and coauthors \cite{MAR13} showed that chimeras emerge naturally from a competition between two antagonistic synchronization patterns in a mechanical experiment involving two subpopulations of identical metronomes
coupled in a hierarchical network. Furthermore, chimeras were experimentally realized using electrochemical oscillators \cite{WIC13} as well as electronic nonlinear delay oscillator \cite{LAR13}.

The importance of chimera states is also very relevant for brain dynamics, since it is believed that they could potentially explain the so-called ``bumps'' of neuronal activity (proposed in mechanisms of visual orientation tuning, the rat head direction system, and working memory \cite{LAI01}) as well as the phenomenon of {\it unihemispheric sleep} \cite{RAT00} observed in dolphins and other animals
which sleep with one eye open, suggesting that one hemisphere of the brain is synchronous the other being asynchronous. 
For this reason, it is particularly interesting that such states were recently observed in leaky integrate-and-fire 
neurons with excitatory coupling \cite{OLM10},  as well as in networks of
FitzHugh-Nagumo \cite{OME13} and Hindmarsh-Rose \cite{HIZ13} oscillators.

Excitability is an important feature of neuronal dynamics \cite{IZH00} as it determines the mechanism of the generation 
of action potentials (spikes)
through which neurons communicate. There are two types of excitability: type I yields a response of finite amplitude and infinite period through
a global bifurcation, and type II gives rise to zero-amplitude and finite period spikes via a Hopf bifurcation.
Type-II excitability is often modeled by the FitzHugh-Nagumo system for which ``multi-chimera'' (or ``clustered 
chimera'' \cite{SET08}) states, which consist of multiple coherent regions, were recently found slightly above the 
excitability threshold \cite{OME13}.
The Hindmarsh-Rose model which is representative for both type-I and type-II excitability, exhibits 
very complex behaviour including spiking, regular and chaotic bursting for which chimera states
and other collective dynamics were identified \cite{HIZ13}.

In this work, we will focus on a generic model for type-I excitability and we will focus on the fundamental
dynamics by performing a systematic analysis as far as chimera states are concerned.
The system under consideration is representative for a global bifurcation, namely a saddle-node bifurcation on a limit cycle
also known as Saddle-Node Infinite PERiod (SNIPER) bifurcation, which is also known as Saddle-Node bifurcation on an 
Invariant Circle (SNIC). It is defined by the following equations \cite{HU93,DIT94,HIZ07,AUS09}:
\begin{equation}\label{eq:SNIPERuncoupled}
\begin{split} 
\dot{x} &= x(1-x^2-y^2)+y(x-b),\\
\dot{y} &= y(1-x^2-y^2)-x(x-b),
\end{split} 
\end{equation}
with the state variables $x(t)$ and $y(t)$, and $b$ is the bifurcation parameter. 
For $b<1$, there are three fixed points: an unstable focus at the origin and a 
pair of a saddle-point and a stable node on the unit circle with coordinates $(b,+\sqrt{1-b^2})$ and $(b,-\sqrt{1-b^2})$, respectively. 
The latter two collide for $b_c=1$ at $(x^*,y^*)=(1,0)$ and a limit cycle with constant radius 
$\rho_c=\sqrt{x^2-y^2}=1$ is born. Above but close to the bifurcation, 
the frequency $f$ of this limit cycle obeys a characteristic square-root scaling law
$f \sim \sqrt{b^2-1}$. 

\begin{figure}[ht!]
  \centering
%   \includegraphics[scale=0.55]{Figure0.pdf}
%   \flushright
  \includegraphics[scale=0.65]{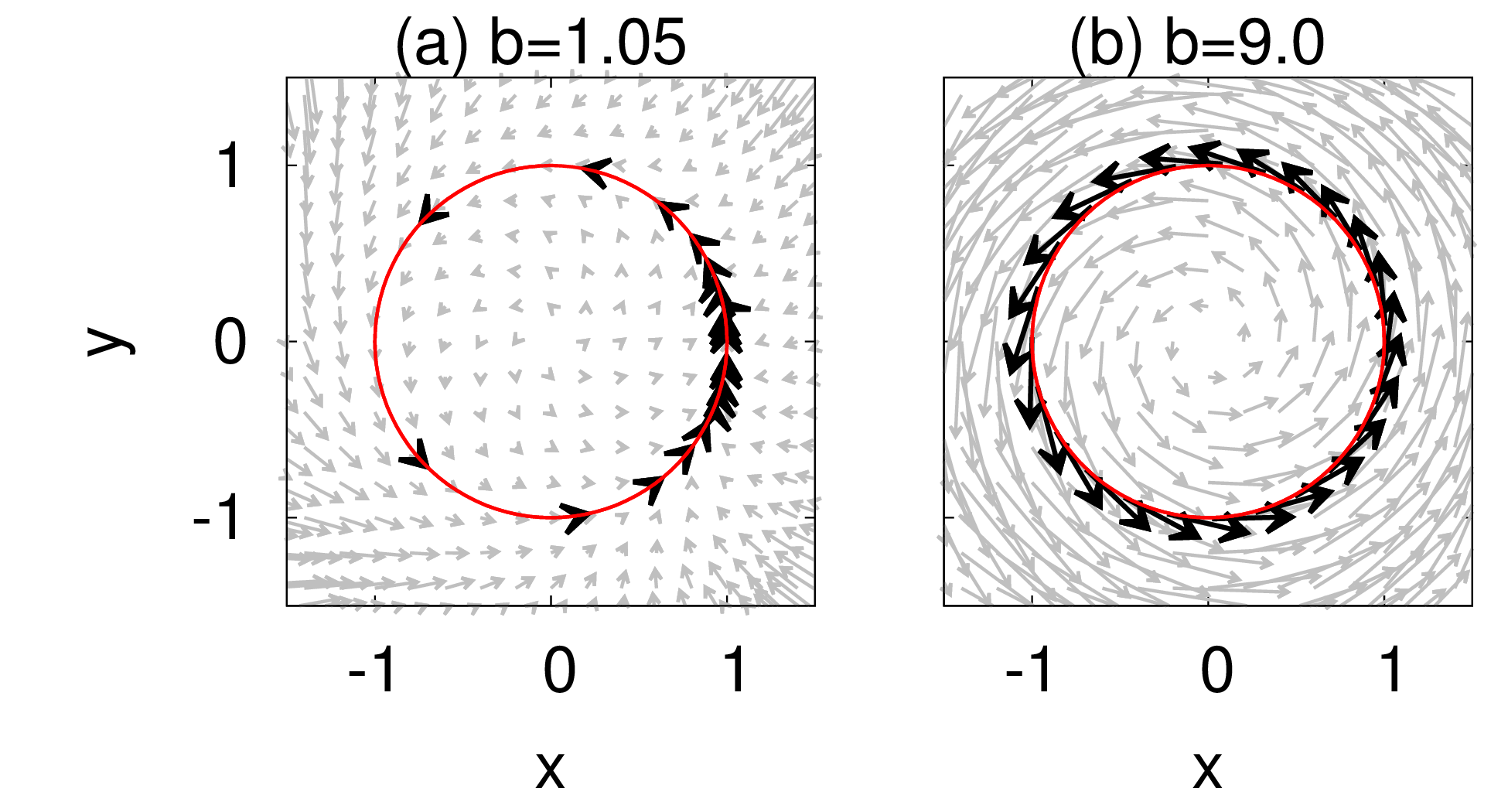}
  \caption{\label{fig:Figure0} SNIPER model in the oscillatory regime: Numerical solution of 
Eqs.~(\ref{eq:SNIPERuncoupled}) for two different values of the bifurcation parameter $b$.
}
\end{figure}

% At the critical value $b_c=1$ of the bifurcation parameter, two saddle-node points collide with each other at $(x_c,y_c)=(1,0)$.
In the following, we choose $b>b_c$ so that the system operates in the oscillatory regime. The system oscillates with 
constant amplitude $\rho=1$ and the period $T_0$ is given by $2\pi/\sqrt{b^2-1}$. In figure~\ref{fig:Figure0} the 
numerical solution of $x$ and $y$ is shown for one period. For $b=1.05$ (figure~\ref{fig:Figure0}(a)), the dense region 
(the so-called ``ghost'') where the system
slows down marks the collision point of the saddle and the node, i.e. $(x^*,y^*)=(1,0)$. For this parameter value,
the system remembers the collision point because it is close to the critical value $b_c$. The phase velocity converges to 
a constant value as soon as $b$ becomes large enough (figure~\ref{fig:Figure0}(b)). 

The rest of this paper is organized as follows: In Sec.~\ref{sec:model}, we introduce the coupling topology and 
describe the main features of the observed dynamics. In Sec.~\ref{sec:impact}, we scan the parameter plane spanned by 
the bifurcation parameter and coupling range. Section~\ref{sec:multistability} focuses on coexistence of chimeras and other 
patterns and in Sec.~\ref{sec:strength}, we address the role of the coupling strength. Finally, we conclude with a 
summary in Sec.~\ref{sec:conclusions}.

\section{The model}
\label{sec:model}
We consider $N$ nonlocally coupled SNIPER oscillators given by Eqs.~(\ref{eq:SNIPERuncoupled}) arranged on a ring:
\begin{equation}\label{eq:SNIPERcoupled}
\begin{split} 
\dot{x}_k=&\;x_k(1-x_k^2-y_k^2)+y_k(x_k-b)+\frac{\sigma}{2R}\sum_{j=k-R}^{k+R}\left[b_{xx}(x_j-x_k)+b_{xy}(y_j-y_k) \right],\\
\dot{y}_k=&\;y_k(1-x_k^2-y_k^2)-x_k(x_k-b)+\frac{\sigma}{2R}\sum_{j=k-R}^{k+R}\left[b_{yx}(x_j-x_k)+b_{yy}(y_j-y_k) \right],      
\end{split}
\end{equation}
where $k = 1,2,\ldots,N$, $\sigma>0$ is the coupling strength, and $R\in[1,N/2]$ is the number of nearest neighbors 
of each oscillator on either side. The limit cases $R=1$ and $R=N/2$ correspond to nearest-neighbour and 
all-to-all coupling, respectively. It is convenient to scale this parameter by the system size, which defines a 
coupling radius $r=R/N\in[1/N,0.5]$.
The coefficients $b_{lm}$, where $l,m\in\{x,y\}$, are given by the elements of the rotational matrix:
\begin{equation}\label{eq:RotMatrix}
\mathbf{B} = \begin{pmatrix} b_{xx} & b_{xy}\\ b_{yx} & b_{yy}\end{pmatrix}=\begin{pmatrix} \cos{\phi} & \sin{\phi}\\ 
-\sin{\phi} & \cos{\phi}\end{pmatrix},
\end{equation} 
where $\phi\in[-\pi,\pi]$. The matrix $\mathbf{B}$ allows for direct $(xx)$- and $(yy)$-coupling as well as 
cross coupling between $x$ and $y$ as in \cite{OME13}.

The diffusive coupling in Eqs.~(\ref{eq:SNIPERcoupled}) is motivated by the electrical synapses (gap junctions) 
linking real neurons.
Neuronal networks have a considerably higher amount of strong short-range connections rather than long-ranged links 
\cite{SCA95,HEL00,HOL03,HEN11}. This property is implemented in our model by means of $R$-nearest-neighbour coupling in 
both directions.
% , where the number and coupling strength of the short-range connections can be controlled by the 
% parameters $R<N/2$ and $\sigma>0$. 
% The long-range connections are neglected in the model. 
Recently, chimera states have also been reported for global coupling involving a mean--field via a nonlinear 
or linear coupling function as well as time delays \cite{SET08,SET13,SCH14a,YEL14}.
The coupling phase $\phi$ parameterizing the matrix $\mathbf{B}$ can be related to the so-called phase lag parameter, which is 
as essential for the existence of chimera states as is the nonlocal coupling \cite{KUR02a,ABR04,OME10a}.

\begin{figure}[ht!]
  \centering
%   \flushright
  \includegraphics[scale=0.5]{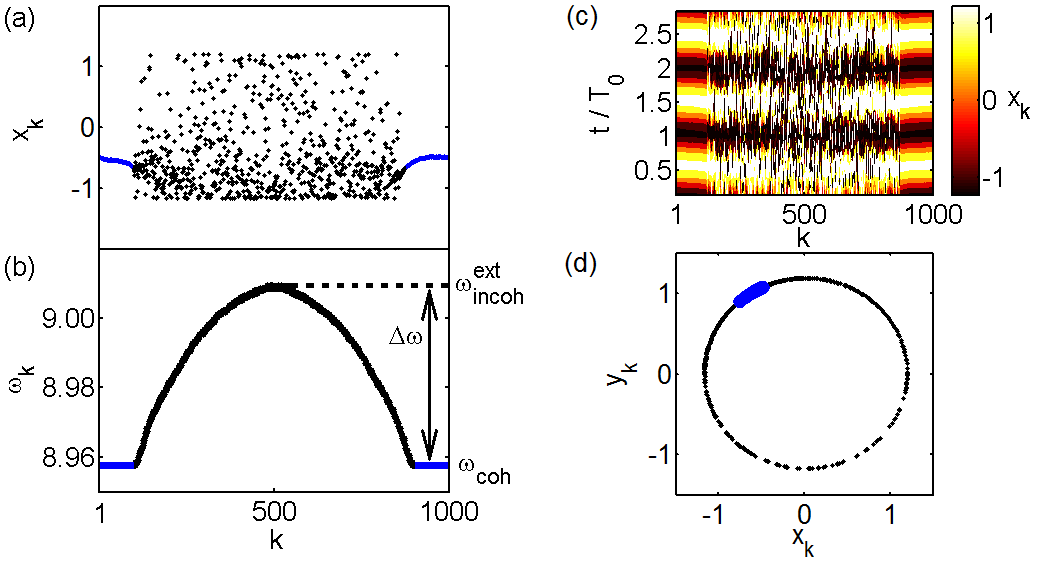}
  \caption{\label{fig:Figure1} Chimera state of nonlocally coupled SNIPER oscillators given by Eqs.~(\ref{eq:SNIPERcoupled}): (a) Snapshot of states $x_k$ and (b) corresponding mean phase velocities $\omega_k$. (c) Space-time plot, where time $t$ is scaled by the period $T_0$ of the uncoupled oscillator. (d) Snapshot in the $(x_k,y_k)$-phase space (blue dots: coherent, black dots: incoherent oscillators). Parameters: $b=9$, $\sigma = 0.1$, $\phi=\pi/2-0.1$, $R=350$, and $N=1000$. For an animation see figure~\ref{fig:animation_b9}.
}
\end{figure}

Figure~\ref{fig:Figure1}(a) shows a snapshot of the variables $x_k$ at a fixed time, providing evidence of a classical chimera state:
One group of neighboring oscillators on the ring is spatially coherent (blue dots) while the remaining elements form a 
a second, spatially incoherent group (black dots).
These two domains of coherent and incoherent oscillators can be distinguished from each other 
through the mean phase velocity of each oscillator $\omega_k=2\pi M_k/\Delta T$, where $M_k$
is the number of periods of the
the $k$th oscillator during a sufficiently long time interval $\Delta T$ \cite{OME13}.

Figure~\ref{fig:Figure1}(b) shows the characteristic profile for the mean phase velocities $\omega_k$ corresponding to the chimera state
of figure~\ref{fig:Figure1}(a). The oscillators in the coherent domain (blue) rotate along the unit circle at a constant speed $\omega_{\text{coh}}$,
whereas the incoherent oscillators (black)
have different mean phase velocities $\omega_{\text{incoh}}$ with a maximum value denoted by $\omega_{\text{incoh}}^{\text{ext}}$.
% are slower than those of the incoherent domain ($\omega_{\text{coh}}<\omega_{\text{incoh}}$).
% The value  denotes the extreme value of the mean phase velocities of the incoherent group. 
If the difference defined as
\begin{equation}\label{eq:DeltaOmega}
\Delta\omega = \omega_{\text{incoh}}^{\text{ext}} - \omega_{\text{coh}},
\end{equation}
is sufficiently larger that a certain threshold value, we can ensure the existence of a chimera state. 
Note that, for the particular chimera state of figure~\ref{fig:Figure1}(a), it holds that $\omega_{\text{incoh}}^{\text{ext}}>\omega_{\text{coh}}$. 
Figure~\ref{fig:Figure1}(c) shows the corresponding space-time plot for the variables $x_k$. For weak coupling, which is the case here, the period of the oscillators
converges to the period $T_0$ of the uncoupled system. Investigations of space-time plots for extended simulation times reveal that the (in)coherent domains
are stationary, i.e. there is no ``drift'' on the ring. Finally, figure~\ref{fig:Figure1}(d) shows the state of each oscillator at a certain time $t$ in phase space (the blue dots mark 
the coherent oscillators while the black dots the incoherent ones).

In the following sections, we will systematically investigate the effect of the bifurcation parameter $b$ as well as the coupling parameters $R$ and $\sigma$ on the chimera state. We will compare our results with findings of previous studies on chimera states in neuronal networks and shed light on new
dynamical features.

For the numerical integration of the Eqs.~(\ref{eq:SNIPERcoupled}) we used the Euler method with step size $dt=0.01$. The initial conditions for $x_k$ and $y_k$ are randomly distributed on the unit circle and we discard transients of $1000$ time steps. For the mean phase velocities $\omega_k$, we average over a time interval $\Delta T=10.000$. 

\section{Impact of the bifurcation parameter and coupling range on the dynamics.}
\label{sec:impact}
\begin{figure}[ht!]
  \centering
  \includegraphics[scale=0.5]{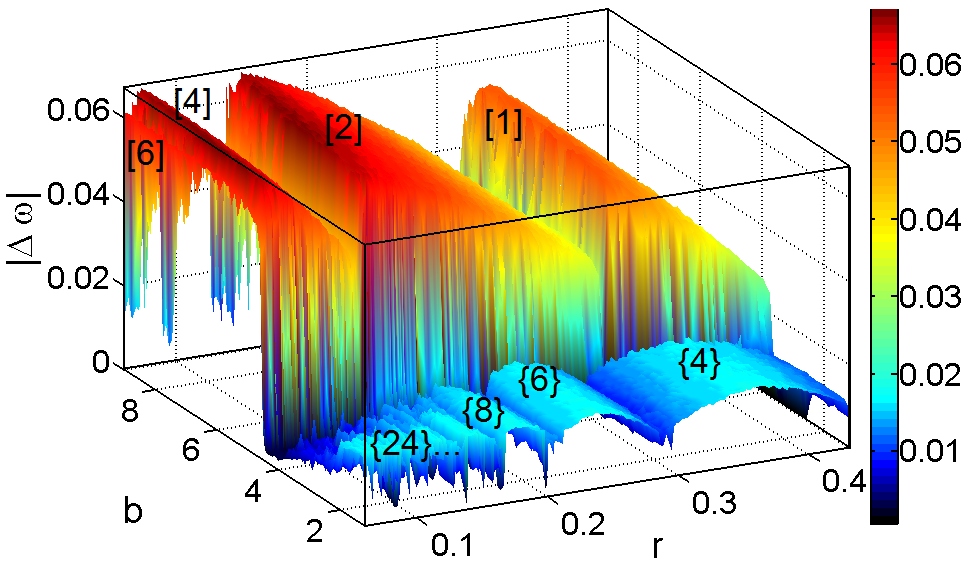}
  \caption{\label{fig:Figure2a} Stability diagram in the $(b,r)$-plane: Modulus of the difference $|\Delta\omega|$ between the mean phase velocities of the coherent and incoherent oscillators (equation~(\ref{eq:DeltaOmega})) as a function of the 
bifurcation parameter $b$ and the coupling radius $r$. The numbers in the brackets
denote the number of the (in)coherent domains of the corresponding chimera state. Square and curly brackets refer to ``normal'' and ``flipped'' $\omega$-profile, respectively. Parameters: $\sigma=0.1$, $\phi=\pi/2-0.1$, and $N=1000$.}
\end{figure}

A stability diagram for the chimera states is displayed in figure~\ref{fig:Figure2a}
where the dependence of the modulus of $\Delta \omega$ (equation~(\ref{eq:DeltaOmega})) is plotted with respect to the bifurcation parameter $b$ and the coupling radius $r= R/N$.

Starting from the values $b=9$, $r=0.43$ and a certain set of initial conditions as described above,
% and a certain set of randomly distributed initial conditions on the unit circle, we compute the states $(x_k,y_k)$ at $t=\Delta T$ and the mean phase velocities $\omega_k$ for each $k\in[1,N]$, where $N=1000$. 
we perform a continuation on the direction of smaller $r$-values down to
% We then
% proceed by letting $r$ decrease from $r=0.43$ 
 $r=0.06$ and calculate $\Delta \omega$ for each coupling radius.
% with a maximum step size of $\Delta r = 0.002$
% and use as the respective initial conditions for each step, the result of each $(x_k,y_k)$ at $t=\Delta T$ of the previous step. 
% In the following, we call this practice `continuation in $r$', when $b$ and $\sigma$ are fixed. 
Subsequently, for values of $r\in[0.04,0.46]$ we perform a continuation in $b$-direction from $b=9$ down to $b=0.1$ starting again at $r=0.43$. 
The coupling strength is fixed at a constant value $\sigma=0.1$.
%  in the interval using 
% the results for all pairs $(x_k,y_k)$ at $t=\Delta T$
%  and each value of
%  as initial conditions for a `continuation in $b$' 
% from $b=9$ to $b=1.2$ 
% in steps of $\Delta b=0.1$.

From figure~\ref{fig:Figure2a} it is clear that $|\Delta \omega|$ has a non-monotonous behaviour in the $(b,r)$-plane.
Each ``bump'' in the 3D surface corresponds to a different type of chimera state associated to a different 
number of (in)coherent domains, marked in the square/curly brackets. Some of these states are explicitly shown below in figure~\ref{fig:Figure2b} for 
certain combinations of $b$ and $r$.

For large values of the bifurcation parameter (red-colored ``bumps'' in figure~\ref{fig:Figure2a} and figure~\ref{fig:Figure2b}(a')) a classical chimera state
with one group of (in)coherent oscillators exists. By decreasing $r$, which physically means removing more and more long-range connections, the number of clustered (in)coherent oscillators increases. In the red-colored ``bumps'' of figure~\ref{fig:Figure2a} these so-called ``multi-chimera'' states 
exhibit the characteristic feature that
 $\omega_{\text{incoh}}^{\text{ext}}>\omega_{\text{coh}}$ (i.e. $\Delta \omega>0$),
shown in the corresponding mean phase velocity profiles in figure~\ref{fig:Figure2b}(b')-(d'). 
We denote  these chimera states, for which $\Delta \omega>0$, by the number of their (in)coherent domains in {\it square} brackets $[1]$, $[2]$, $[4]$, and $[6]$.

\begin{figure}[ht!]
  \centering
  \includegraphics[scale=0.38]{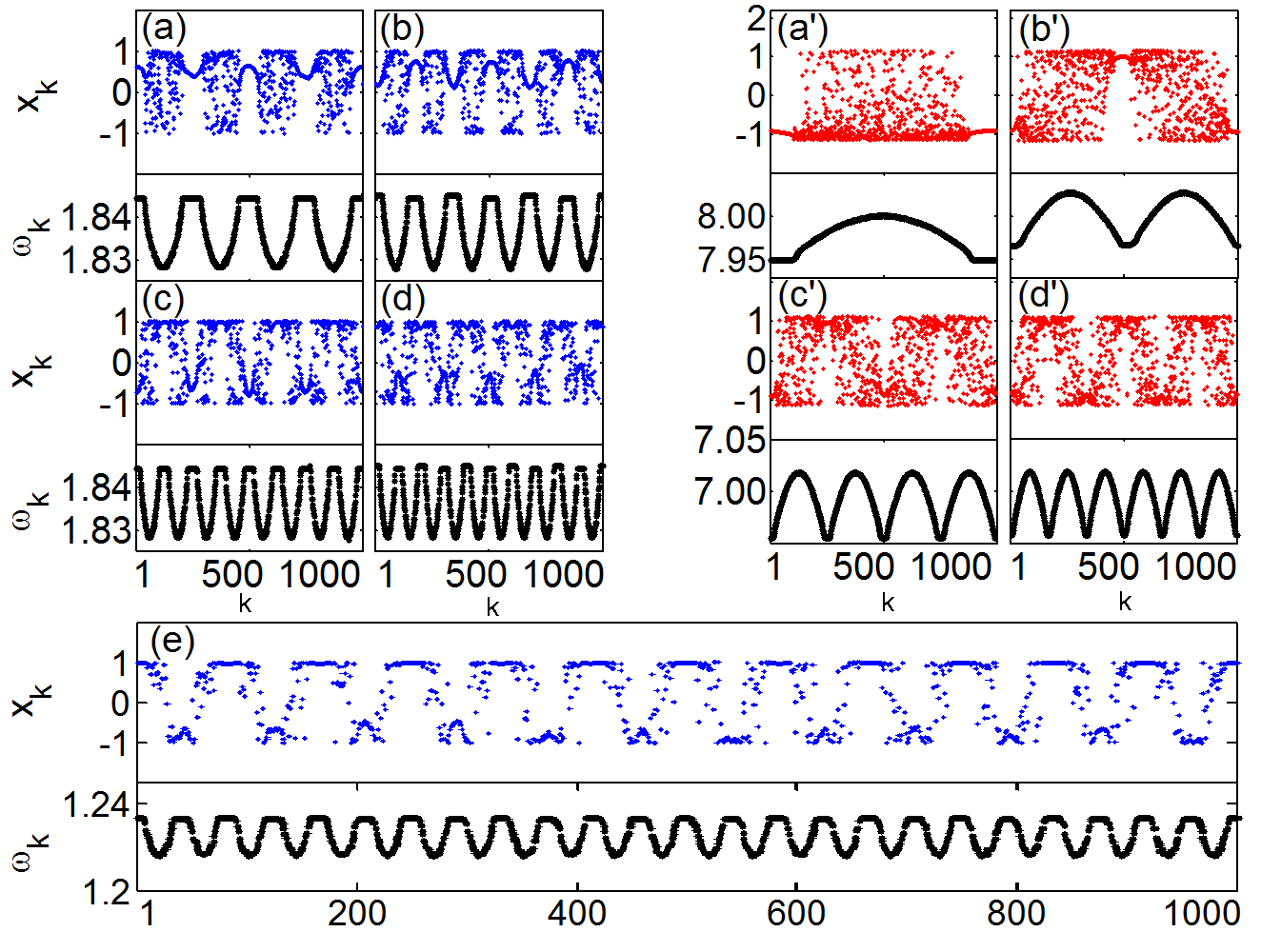}
  \caption{\label{fig:Figure2b} Clustered chimera states of figure~\ref{fig:Figure2a}: Snapshots  of the states $x_k$ at 
different points in the $(b,r)$-plane. Red and blue dots correspond to ``normal'' and ``flipped'' $\omega$-profiles 
(black dots), respectively. (a) $r=0.35$, $b=2$ (for an animation see figure~\ref{fig:animation_b2}), (b) $r=0.24$, $b=2$, (c) $r=0.18$, $b=2$, (d) $r=0.14$, $b=2$, (e) 
$r=0.06$, $b=2$, (a') $r=0.35$, $b=8$, (b') $r=0.18$, $b=8$, (c') $r=0.08$, $b=7$, (d') $r=0.06$, $b=7$.
Other parameters: $\sigma = 0.1$, $\phi=\pi/2-0.1$, $N=1000$.}
\end{figure}

For lower values of $b$ (blue-colored ``bumps'' in figure~\ref{fig:Figure2a} and figures~\ref{fig:Figure2b}(a-e)), we exclusively find multi-chimera states. As in the case of larger $b$, the number of clustered (in)coherent oscillators increases with decreasing coupling radius $r$. 
However, there is a significant difference:  The mean phase velocities of the incoherent oscillators is smaller than 
the velocity of the coherent ones, i.e. $\Delta \omega<0$. 
Hence, there exists a critical value of the bifurcation parameter (found to be around $b=4$), where $\Delta \omega$ changes its sign, resulting in a ``flip'' in the mean phase
velocity profile.
The chimera states with a ``flipped'' $\omega$-profile are 
denoted by the number of (in)coherent domains in {\it curly} brackets $\{4\}$, $\{6\}$, $\{8\}$, $\{10\}$, \dots, $\{24\}$.
% 
% by the number of their (in)coherent domains in {\emph curly} brackets $\{2\}$-, $\{4\}$-, $\{6\}$-.
% 
% % and thus we have that $\omega_{\text{incoh}}^{\text{ext}}<\omega_{\text{coh}}$.
% 
%  We will revert to this phenomenon in another paragraph.

% But these -chimeras (blue) differ from the $\{1\}$-, $\{2\}$-, $\{4\}$-, $\{6\}$-chimeras (red).

% \textcolor{blue}{---Mainly even multi-chimeras (?):---} \textcolor{red}{AV: Hier sollte noch was zu den geraden/ungeraden Chimeras hin. Warum gibt es hauptsächlich gerade Chimeras?}

The characteristic form of the average phase velocities profile is commonly considered as a criteria to distinguish chimera states in the systems of coupled oscillators. The most often observed in the variety of systems is the case when the coherent oscillators perform smaller average phase frequencies, and incoherent oscillators are faster. However, the opposite situation is also possible, when the coherent oscillators perform faster oscillations as the incoherent ones. In the system of nonlocally delay coupled phase oscillators, two types of chimera states were distinguished  depending on whether the effective frequencies of the incoherent oscillators are larger or smaller than the frequencies of the coherent ones \cite{OME08,OME10b}. The regions of stability for these two types of chimera states depend on the time delay and strength of the coupling. Moreover, both types of chimera states can coexist.

The ``flipped'' phase velocities profile was also observed in systems which do not consider time delay in the coupling, but has not been explained so far. The Kuramoto model with repulsive coupling allows for multi-chimera states for which the phase velocity profiles show larger average frequencies for oscillators that belong to coherent domain~\cite{MAI14}. Similar behaviour is also observed for chimera states with one incoherent domain in the complex Ginzburg-Landau equation with nonlocal coupling \cite{SET13}.
%  with strong coupling limit, where amplitude variations must be taken into account, 
% also exhibits similar 
In that system, however, chimera states with multiple incoherent domains possess the usually observed mean phase velocity profiles.

Hence, the flip of the phase velocities can not be explained only by the influence of time delay, or strong coupling. This feature was observed in experiments as well, in networks of electrochemical oscillators with nonlocal coupling, the frequencies of the oscillators from the coherent domain of chimera state are larger than the frequencies of the incoherent ones~\cite{WIC13}. 

In our system, we observe direct dependence of the form of the mean phase velocity profile on the parameter $b$ defining the frequency of the local uncoupled unit.

\section{Multistability of patterns: Coexisting chimeras and traveling waves.}
\label{sec:multistability}
\begin{figure}[ht!]
  \centering
  \includegraphics[scale=0.3]{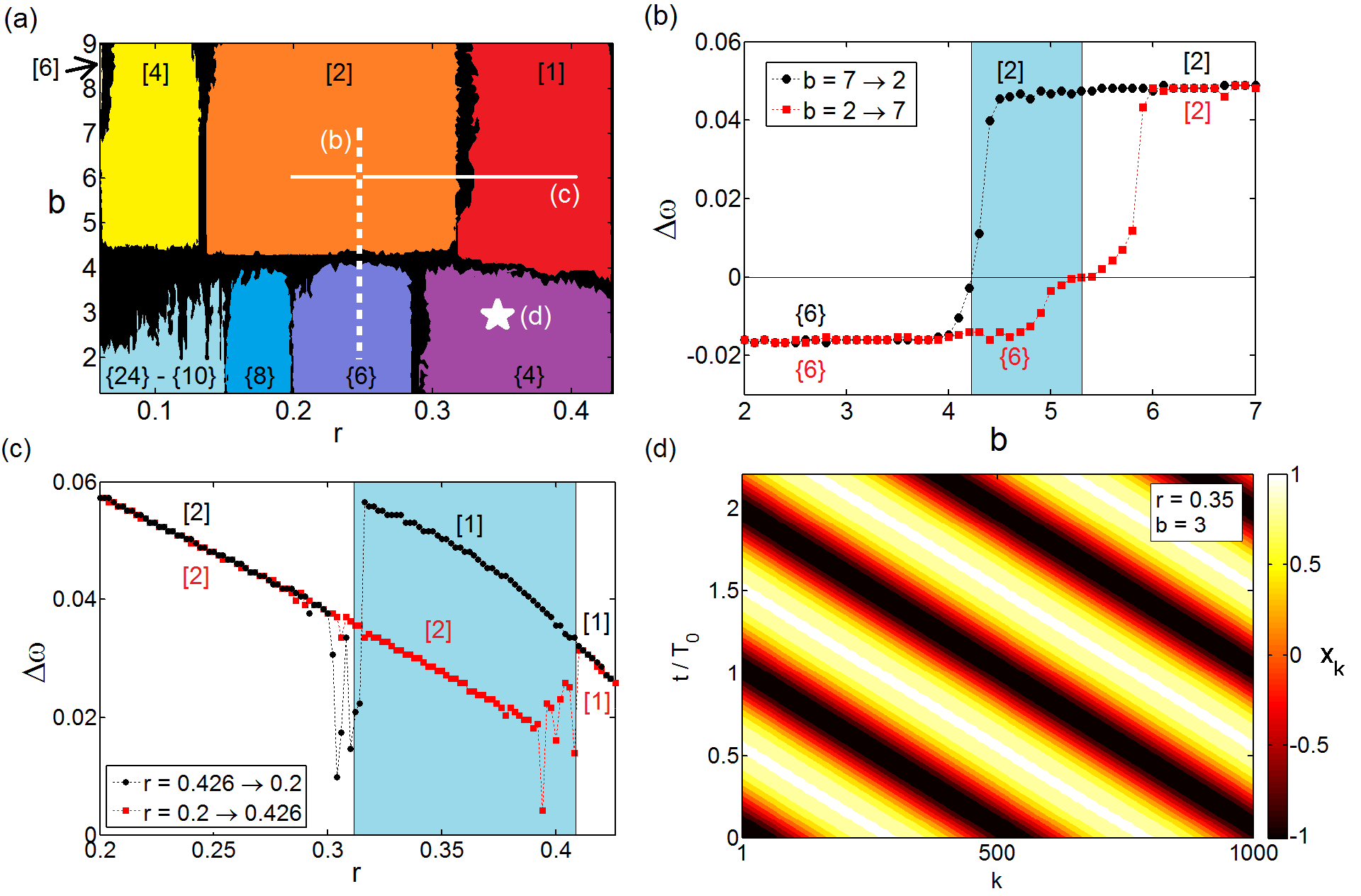}
  \caption{\label{fig:Figure3} Coexisting chimera states and traveling waves: (a) projection to the $(b,r)$-plane of figure~\ref{fig:Figure2a}. 
% The white lines and star denote continuation in both parameters fixed system parameters which are used in the following panels: 
(b) Up and down sweep in $b$-direction as marked by the dashed white line in figure~\ref{fig:Figure3}(a) for fixed $r=0.25$. $[2]$- and $\{6\}$-chimera states coexist in the shaded (light-blue) area. (c) Up and down sweep in $r$-direction as marked by the solid white line in figure~\ref{fig:Figure3}(a) for fixed $b=6$. $[1]$- and $[2]$-chimera states coexist in the shaded (light-blue) area. (d) Traveling wave solution, which coexists with the $\{4\}$-chimera, for $r$ and $b$ marked by the white star in figure~\ref{fig:Figure3}(a). The time is scaled by the period $T_0$ of an uncoupled oscillator. Other Parameters: $\sigma = 0.1$, $\phi=\pi/2-0.1$, and $N=1000$.
}
\end{figure}

% \textcolor{blue}{---Intro multistable solutions:---} 
The coexistence of different multi-chimeras, traveling waves, and completely synchronized states in the phase space has been observed in many other systems 
of nonlocal coupled oscillators \cite{OME13,HIZ13,DZI13}. Depending on the initial conditions the stationary state can vary significantly. Such multistable solutions are also possible in system~(\ref{eq:SNIPERcoupled}) as demonstrated in figure~\ref{fig:Figure3}.

A schematic representation of the identified multi-chimeras in the $(b,r)$-plane is shown in figure~\ref{fig:Figure3}(a) .
Each region has a different colour associated to a different chimera type as described in the previous section.   
The black regions correspond to intermittent states, which are mainly desynchronized. Along the white lines, figures~\ref{fig:Figure3}(b) and (c)
display the results of a continuation in $b$ (dashed line) and $r$ (solid line), respectively. The continuation is performed as down sweep in $b$ (or $r$) and then repeated in the opposite direction. In both cases we find a region where different types of chimera states coexist.

In particular, for intermediate values of the bifurcation parameter $b$, there is coexistence of a $[2]$- and 
$\{6\}$-chimera state marked by the shaded (light blue) area of figure~\ref{fig:Figure3}(b). This area of coexisting 
chimera states, moreover, marks the transition between ``flipped'' ($\Delta \omega<0$) and ``normal'' ($\Delta \omega>0$) mean
phase velocity profile. This transition occurs at a different and, in particular, lower value of $b$ when the continuation is performed 
in the direction of decreasing $b$ (black dots) 
than when performed in the opposite direction (red squares), i.e. our system exhibits, apart from multistability, hysteresis phenomena as well.   

%  	
% We did many other numerical experiments and we could show that for the parameter set given in figure~\ref{fig:Figure3} the neighboring chimera states all overlap with each other. 
Coexisting chimera states may also be found by varying parameter $r$, as shown in figure~\ref{fig:Figure3}(c): Depending on the choice
of initial conditions, one may observe either a $[1]$- or a $[2]$-chimera state (shaded, light-blue area) both with $\Delta \omega>0$.
In both increasing (red dots) and decreasing $r$ (black dots) directions, there are deviations from the piecewise linear 
behaviour of $\Delta \omega (r)$ 
which correspond to desynchronized states.

The observed multi-chimera states may also coexist with completely synchronized states and traveling waves. One example of such a point in parameter space 
is marked by the white star in figure~\ref{fig:Figure3}(a) and the corresponding space-time pattern is shown in figure~\ref{fig:Figure3}(d). This is a traveling wave solution of wave number $2$ coexisting with a $\{4\}$-chimera state. The time in the vertical axis is scaled by the period $T_0$ of the uncoupled oscillator. Multistability between traveling waves and breathing states have recently alse been reported for chaotic systems with nonlocal coupling \cite{DZI13}.

\section{Role of the coupling strength.}
\label{sec:strength}
In order to complete our study on the effect of the system parameters on the
dynamics of chimera states, we will investigate the role of the 
coupling strength $\sigma$ in this section. 

Again, we perform a parameter continuation and 
focus on the behaviour of $\Delta \omega$ as $\sigma$ increases for   
different multi-chimera states. Our findings show that even at large $\sigma$ the
corresponding multi-chimera state is preserved. However, we observe that, for 
certain values of the bifurcation parameter $b$ and the coupling radius $r$, the coupling
strength may induce a spatial motion of domains of the (in)coherent oscillators.

\begin{figure}[ht!]
  \centering
%   \flushright
  \includegraphics[scale=0.39]{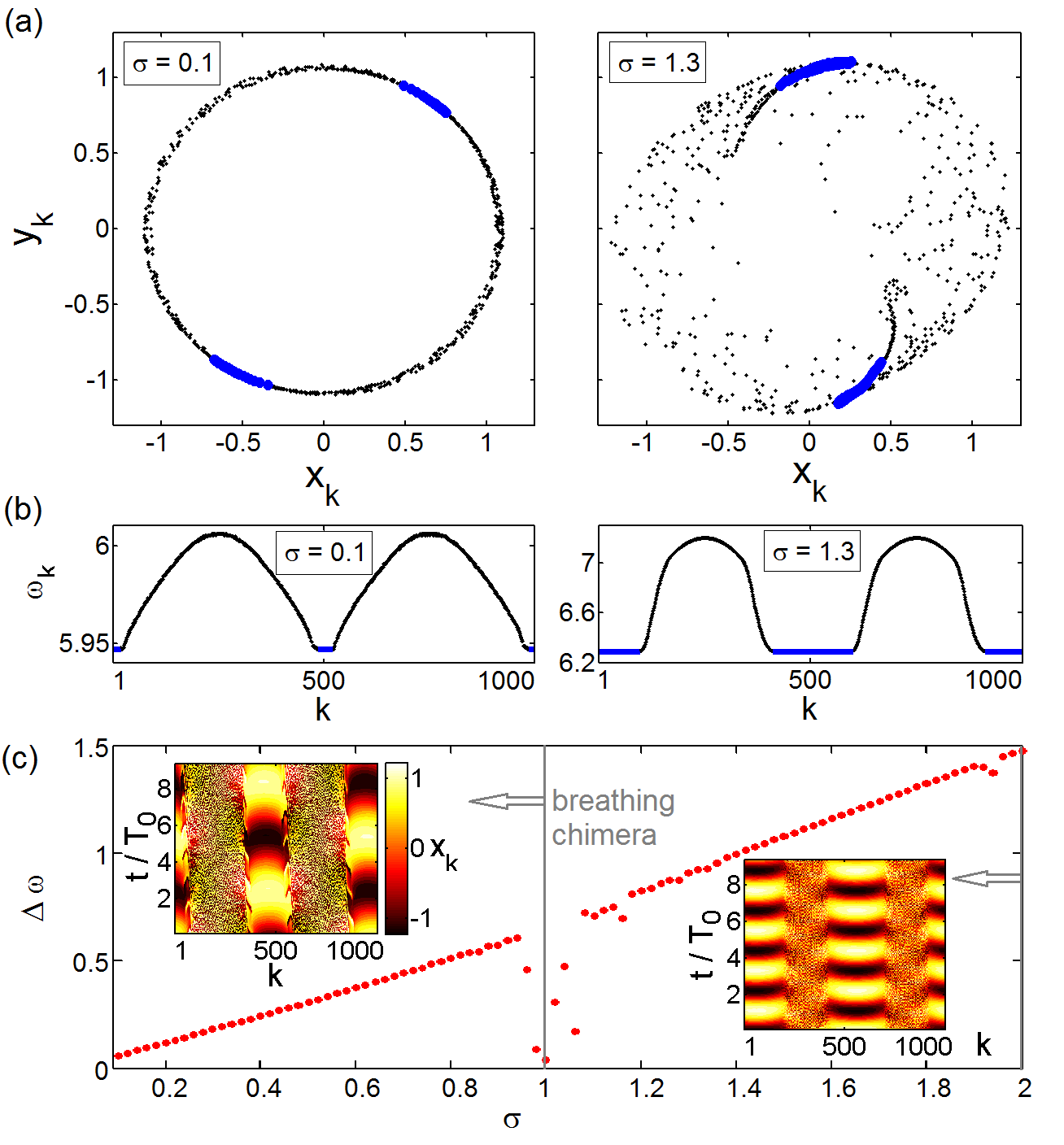}
  \caption{\label{fig:Figure4} Impact of the coupling strength on the
$[2]$-chimera state: (a) Snapshots in the $(x_k,y_k)$-plane for
different coupling strengths $\sigma$. (b) Corresponding mean phase
velocities $\omega_k$ and (c) $\Delta \omega$ as a function
of the coupling strength. The insets are a space-time plot for a fixed
$\sigma$ (left inset: $\sigma = 1.0$, right inset: $\sigma = 2.0$).
Other parameters: $b=6$, $\phi=\pi/2-0.1$, $R=190$, and $N=1000$.
}
\end{figure}

Figure~\ref{fig:Figure4} shows the results for the
$[2]$-chimera state associated with the orange regime of figure~\ref{fig:Figure3}(a).
With increasing coupling strength, each oscillator becomes more and more
influenced by the dynamics of the remaining oscillators. Therefore, the
trajectories of the incoherent oscillators, in particular, begin to deviate significantly from the unit circle as shown in
the right plot of figure~\ref{fig:Figure4}(a) for $\sigma=1.3$. 

The corresponding mean phase velocity profiles $\omega_k$ of the $[2]$-chimera
state can be seen in figure~\ref{fig:Figure4}(b). 
For larger $\sigma$ (right plot) the total number of coherent oscillators increases
while the number of incoherent oscillators decreases. Figure~\ref{fig:Figure4}(c)
shows  that the difference between the mean phase velocity of the
coherent and incoherent oscillators $\Delta \omega$ linearly increases
with the coupling strength, apart for a narrow range of $\sigma\approx 1$ where $\Delta \omega$ deviates. 
In this regime, the corresponding
space-time plots of the $[2]$-chimera state reveal that the
(in)coherent domains start to move spatially with time (see left inset of figure~\ref{fig:Figure4}(c)).  Beyond
this regime of  moving patterns, the $[2]$-chimera state is stationary (see right inset
of figure~\ref{fig:Figure4}(c)).
% For weak coupling the period of the
% coherent oscillators is close to the period $T_0$ of the uncoupled
% oscillator.
Comparing the two insets of figure~\ref{fig:Figure4}(c),
we observe that for increasing coupling strength the period of the coherent oscillators increases.
On the other hand, the period strongly decelerates once the
$[2]$-chimera state introduces spatial motion.

In general, chimera states can be stationary or can perform two types of motion in space, in which the coherent and incoherent domains change their spatial position in time. The first one is a chaotic motion of the position of the chimera observed in nonlocally coupled phase oscillators. Such a motion shows a sensitive dependence on the initial conditions and is a finite-size effect that vanishes in the thermodynamic limit. It can be described as a Brownian motion and depends on the coupling radius, the phase lag parameter, and the shape of the coupling function~\cite{OME10a}.
The second type is a periodic motion of the coherent and incoherent domains of the chimera state, called ``breathing chimera''.  Breathing chimeras were first observed in the system of two oscillator populations in which each oscillator is coupled equally to all the others in its group, and less strongly to those in the other group~\cite{ABR08}, and recently in the nonlocal complex Ginzburg-Landau equation with strong coupling limit~\cite{SET13}.

The numerical evidence shows, that spatial motion of coherent and incoherent domains in our system is periodic, thus we conclude that for distinct  values  of parameter~$b$ we observe the phenomenon of breathing chimera in our system.

% \section{Phase reduction}
% 
% If the coupling strength $\sigma$ is small we know from the numerics that the amplitude $r_k=\sqrt{x_k^2+y_k^2} \approx 1\;\forall k$. Thus in polar representation $(r_k,\varphi_k)$ the two equations~(\ref{eq:SNIPERcoupled}) are reducible to one phase equation. The transformation formulas read $x_k=r_k\cos\varphi_k$ and $y_k=r_k\sin\varphi_k$, where $r_k\equiv 1\;\forall k$ and $\varphi_k$ is the phase. After transformation of the equations~(\ref{eq:SNIPERcoupled}) and under the condition that $\sigma << 1$ the corresponding phase equation is given by 
% \begin{equation}\label{eq:reducedSNIPER(1)}
% \dot{\varphi}_k = b - \cos\varphi_k + \sigma \sin\phi + \frac{\sigma}{2R}\sum_{j=k-R}^{k+R} \sin\left(\varphi_j - \varphi_k-\phi \right).
% \end{equation}
% With respect to $b>>1$ the terms $\cos\varphi_k\in[-1,1]$ and $\sin\phi\in[-1,1]$ can be neglected and the phase equation~(\ref{eq:reducedSNIPER(1)}) can be rewritten as
% \begin{equation}\label{eq:reducedSNIPER(2)}
% \dot{\varphi}_k = b + \frac{\sigma}{2R}\sum_{j=k-R}^{k+R} \sin\left(\varphi_j - \varphi_k-\phi \right).
% \end{equation}
% 
% 
% 
% \vfill

\section{Conclusions}
\label{sec:conclusions}

In this work, we have verified the occurrence of clustered chimera states in 
a generic model for a saddle-node bifurcation on a limit cycle 
representative for neural excitability type-I.
This, along with recent reports on multi-chimera states in
nonlocally coupled FitzHugh-Nagumo  \cite{OME13} and Hindmarsh-Rose \cite{HIZ13} oscillators
provide strong evidence that this kind of symmetry breaking is very relevant for applications in neuroscience. 

In particular, we presented a detailed exploration of the parameter space, where chimera states occur, and 
investigate the dependence on the proximity to the excitability
threshold and the range of the nonlocal coupling. We identified chimera states for which the mean phase velocity
has a ``flipped'' profile. A similar result was also reported in a recent study of 
Kuramoto oscillators with repulsive coupling \cite{MAI14}.
Findings of coexisting chimera states and traveling waves
in the parameter space establish the existence of multistability in our model.
Finally, it was shown that for increasing coupling strength the domains of 
coherent oscillators become bigger and at the same time spatial motion
of the incoherent oscillators is observed.

 \ack
We thank A. Provata and E. Sch\"oll for stimulating discussions. 
This work was supported by the German Academic Exchange Service \textit{DAAD} and the Greek State Scholarship Foundation
\textit{IKY} within the PPP-IKYDA framework.
IO and PH acknowledge support by BMBF (grant no. 01Q1001B) in the framework of BCCN  Berlin (Project A13). AV and IO acknowledge support by DFG in the framework of the Collaborative Research Center 910.  This research has been co--financed by the European Union (European Social Fund--ESF) 
and Greek national funds through the Operational Program ``Education and Lifelong Learning'' of the National Strategic Reference 
Framework (NSRF)--Research Funding Program: Thales. Investing in knowledge society through the European Social Fund.

\section*{References}

% \bibliographystyle{iopart-num}
% \bibliographystyle{unsrt}
% \bibliographystyle{prsty-fullauthor}
% \bibliography{ref}

\begin{thebibliography}{10}
\expandafter\ifx\csname url\endcsname\relax
  \def\url#1{{\tt #1}}\fi
\expandafter\ifx\csname urlprefix\endcsname\relax\def\urlprefix{URL }\fi
\providecommand{\eprint}[2][]{\url{#2}}
% Bibliography created with iopart-num v2.1
% /biblio/bibtex/contrib/iopart-num

\bibitem{KUR02a}
Kuramoto Y and Battogtokh D 2002 {\em Nonlin. Phen. in Complex Sys.\/} {\bf 5}
  380

\bibitem{ABR04}
Abrams D~M and Strogatz S~H 2004 {\em Phys.~Rev.~Lett.\/} {\bf 93} 174102

\bibitem{PAN14}
Panaggio M~J and Abrams D~M 2014 {\em arXiv:1403.6204\/}

\bibitem{ABR08}
Abrams D~M, Mirollo R, Strogatz S~H and Wiley D~A 2008 {\em Phys. Rev. Lett.\/}
  {\bf 101} 084103

\bibitem{LAI10}
Laing C~R 2010 {\em Phys. Rev. E\/} {\bf 81} 066221

\bibitem{BOU14}
Bountis T, Kanas V, Hizanidis J and Bezerianos A 2014 {\em arXiv:1308.5528\/}

\bibitem{SET13}
Sethia G~C, Sen A and Johnston G~L 2013 {\em Phys. Rev. E\/} {\bf 88} 042917

\bibitem{ZAK14}
Zakharova A, Kapeller M and Sch{\"o}ll E 2014 {\em Phys.~Rev.~Lett.\/} {\bf
  112} 154101

\bibitem{OME11}
Omelchenko I, Maistrenko Y~L, H{\"o}vel P and Sch{\"o}ll E 2011 {\em Phys. Rev.
  Lett.\/} {\bf 106} 234102

\bibitem{OME12}
Omelchenko I, Riemenschneider B, H{\"o}vel P, Maistrenko Y~L and Sch{\"o}ll E
  2012 {\em Phys. Rev.~E\/} {\bf 85} 026212

\bibitem{OME13a}
Omel'chenko O~E 2013 {\em Nonlinearity\/} {\bf 26} 2469

\bibitem{TIN12}
Tinsley M~R, Nkomo S and Showalter K 2012 {\em Nature Physics\/} {\bf 8}
  662

\bibitem{HAG12}
Hagerstrom A~M, Murphy T~E, Roy R, H{\"o}vel P, Omelchenko I and Sch{\"o}ll E
  2012 {\em Nature Physics\/} {\bf 8} 658

\bibitem{MAR13}
Martens E~A, Thutupalli S, Fourri{\`e}re A and Hallatschek O 2013 {\em Proc.
  Nat. Acad. Sciences\/} {\bf 110} 10563

\bibitem{WIC13}
Wickramasinghe M and Kiss I~Z 2013 {\em PLoS ONE\/} {\bf 8} e80586

\bibitem{LAR13}
Larger L, Penkovsky B and Maistrenko Y~L 2013 {\em Phys. Rev. Lett.\/} {\bf
  111} 054103

\bibitem{LAI01}
Laing C~R and Chow C~C 2001 {\em Neural Computation\/} {\bf 13} 1473

\bibitem{RAT00}
Rattenborg N~C, Amlaner C~J and Lima S~L 2000 {\em Neurosci. Biobehav. Rev.\/}
  {\bf 24} 817

\bibitem{OLM10}
Olmi S, Politi A and Torcini A 2010 {\em Europhys. Lett.\/} {\bf 92} 60007

\bibitem{OME13}
Omelchenko I, Omel'chenko O~E, H{\"o}vel P and Sch{\"o}ll E 2013 {\em Phys.
  Rev. Lett.\/} {\bf 110} 224101

\bibitem{HIZ13}
Hizanidis J, Kanas V, Bezerianos A and Bountis T 2014 {\em Int. J.~Bifurcat.
  Chaos\/} {\bf 24} 1450030

\bibitem{IZH00}
Izhikevich E~M 2000 {\em Int. J.~Bifurcat. Chaos\/} {\bf 10} 1171

\bibitem{SET08}
Sethia G~C, Sen A and Atay F~M 2008 {\em Phys.~Rev.~Lett.\/} {\bf 100} 144102

\bibitem{HU93}
Hu B~Y~K and Das~Sarma S 1993 {\em Phys.~Rev.~B\/} {\bf 48} 5469

\bibitem{DIT94}
Ditzinger T, Ning C~Z and Hu G 1994 {\em Phys.~Rev.~E\/} {\bf 50} 3508

\bibitem{HIZ07}
Hizanidis J, Aust R and Sch{\"o}ll E 2008 {\em Int.~J.~Bifur.~Chaos\/} {\bf 18}
  1759

\bibitem{AUS09}
Aust R, H{\"o}vel P, Hizanidis J and Sch{\"o}ll E 2010 {\em Eur. Phys.~J.~ST\/}
  {\bf 187} 77

\bibitem{SCA95}
Scannell J~W, Blakemore C and Young M~P 1995 {\em J.~Neurosci.\/} {\bf 15(2)}
  1463

\bibitem{HEL00}
Hellwig B 2000 {\em Biol.~Cybern.\/} {\bf 82} 111

\bibitem{HOL03}
Holmgren C, Harkany T, Svennenfors B and Zilberter Y 2003 {\em J.~Physiol.\/}
  {\bf 551.1} 139

\bibitem{HEN11}
Henderson J~A and Robinson P~A 2011 {\em Phys. Rev. Lett.\/} {\bf 107}(1)
  018102

\bibitem{SCH14a}
Schmidt L, Sch{\"o}nleber K, Krischer K and Garcia-Morales V 2014 {\em Chaos\/}
  {\bf 24} 013102

\bibitem{YEL14}
Yeldesbay A, Pikovsky A and Rosenblum M 2014 {\em Phys. Rev. Lett.\/} {\bf
  112}(14) 144103

\bibitem{OME10a}
Omel'chenko O~E, Wolfrum M and Maistrenko Y~L 2010 {\em Phys. Rev.~E\/} {\bf
  81} 065201(R)

\bibitem{OME08}
Omel'chenko O~E, Maistrenko Y~L and Tass P~A 2008 {\em Phys. Rev. Lett.\/} {\bf
  100} 044105

\bibitem{OME10b}
Omel'chenko O~E, Maistrenko Y~L and Tass P~A 2010 {\em Phys. Rev. E\/} {\bf 82}
  066201

\bibitem{MAI14}
Maistrenko Y~L, Vasylenko A, Sudakov O, Levchenko R and Maistrenko V~L 2014
  {\em arXiv: 1402.1363v1\/}

\bibitem{DZI13}
Dziubak V, Maistrenko Y~L and Sch{\"o}ll E 2013 {\em Phys. Rev.~E\/} {\bf 87}
  032907

\end{thebibliography}
\providecommand{\newblock}{}

\clearpage
\appendix

% \section*{Appendix}
\section{Animations}

\begin{figure}[ht!]
\centering
\includegraphics[width=0.7\textwidth]{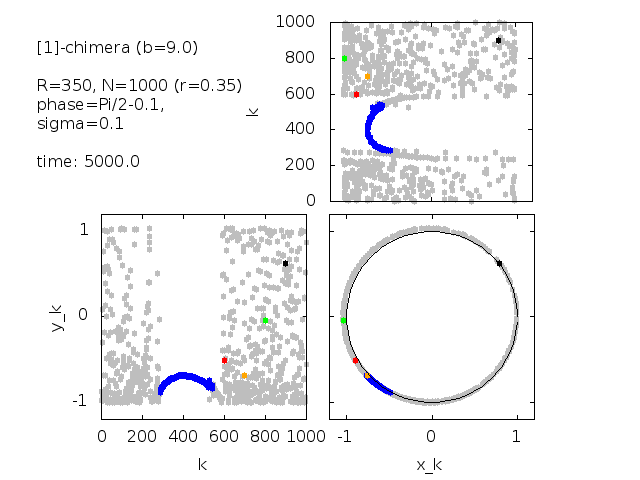}
\caption{\label{fig:animation_b9} Animation of time series in figure~\ref{fig:Figure1} in the time interval $t\in[5000,5020]$. Parameters $b=9$, $\phi=\pi/2-0.1$, $R=350$, and $N=1000$.
}
\end{figure}

\begin{figure}[ht!]
\centering
\includegraphics[width=0.7\textwidth]{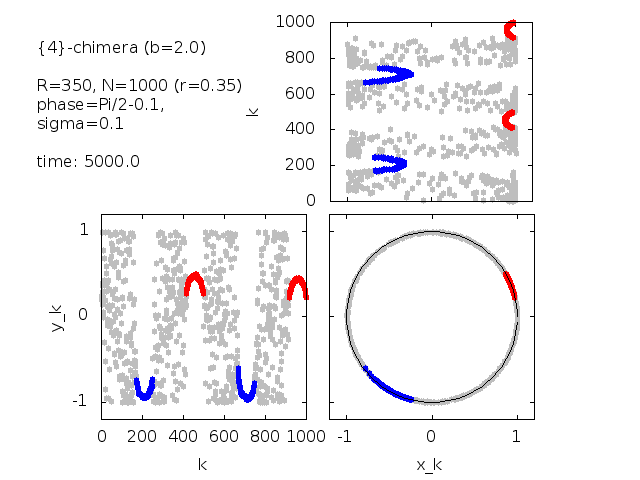}
\caption{\label{fig:animation_b2} Animation of time series in figure~\ref{fig:Figure2b}(a) in the time interval $t\in[5000,5020]$. Parameters $b=2$, $\phi=\pi/2-0.1$, $R=350$, and $N=1000$.
}
\end{figure}

\end{document}